\documentclass[aps,prd,twocolumn]{revtex4}
\usepackage{amsmath,amssymb}

\def\C{\mathbb{C}}
\def\tr{\mathrm{tr}}
\newcommand{\eqa}{\begin{eqnarray}}
\newcommand{\neqa}{\end{eqnarray}}
\newcommand{\be}{\begin{equation}}
\newcommand{\ee}{\end{equation}}

\newcommand{\Ref}[1]{(\ref{#1})}
\newcommand{\scrM}{\mathcal{M}}
\newcommand{\dual}{{}^\star\!}
\newcommand{\Hil}{\mathcal{H}}
\newcommand{\Link}{\mathrm{Link}}

\renewcommand{\texttt}{{}}

\begin{document}

\title{LQG vertex with finite Immirzi parameter}

\author{Jonathan Engle${}^a$, Etera Livine${}^b$, Roberto Pereira${}^a$, Carlo Rovelli${}^a$}
\affiliation{
${}^a$Centre de Physique Th\'eorique de Luminy, Case 907, F-13288 Marseille,
EU\\
${}^b$Laboratoire de Physique, ENS Lyon, CNRS UMR 5672, 46 All\'ee d'Italie, 69364
Lyon, EU}
\date{\small\today}

\begin{abstract}\noindent
We extend the definition of the ``flipped" loop-quantum-gravity vertex to the case of a finite
Immirzi parameter $\gamma$. We cover both the euclidean and lorentzian cases.  We show that the
resulting dynamics is defined on a Hilbert space isomorphic to the one of loop quantum gravity, and
that the area operator has the same discrete spectrum as in loop quantum gravity. This includes the
correct dependence on $\gamma$, and, remarkably, holds in the lorentzian case as well. The
\emph{ad hoc} flip of the symplectic structure that was required to derive the flipped vertex is
not anymore required for finite $\gamma$.  These results establish a bridge between canonical loop
quantum gravity and the spinfoam formalism in four dimensions.
\end{abstract}

\maketitle

\section{Introduction}

The Barrett-Crane (BC) vertex, which provides a tentative definition of
the quantum-gravity dynamics, has been extensively investigated
during the past years  \cite{BC}; its amplitude is essentially given by a Wigner $10j$
symbol.   A different vertex has been recently  introduced in
\cite{eprlett,eprpap}; its amplitude essentially given by the
square of an $SU(2)$ Wigner $15j$ symbol.  There are indications that this new vertex could
ameliorate the properties of the BC model.  First, it appears to correct an over-imposition of the
constraints that was remarked in the derivation of the BC vertex. Second, it does not appear to
freeze the angular degrees of freedom of the gravitational field (that is, $g_{ab}(x)$ for $a\ne
b$) as it has been argued the BC model might do \cite{ema}. Third, preliminary numerical
investigations appear to be consistent with the expectation that geometry wave packets are
propagated by this new vertex in a way consistent with euclidean general relativity (GR)
\cite{numer}. And finally, its kinematics matches exactly the one of the canonical
quantization of GR, as given by loop quantum gravity (LQG) \cite{lqg}.
The vertex was defined in \cite{eprlett,eprpap} only for the euclidean case,
and in the absence of an Immirzi parameter $\gamma$.

A key step to extend the definition of this new vertex was taken in
\cite{L}, where a lorentzian version of the vertex amplitude is constructed, still
without $\gamma$.   Here we extend the construction of the vertex
to the general case of finite $\gamma$, both for the euclidean and the
lorentzian sectors.

As long emphasized by Sergei Alexandrov \cite{sergei2007}, the key technical problem
is how to impose the second class quantum constraints in a covariant way (see
 \cite{covariant}).  These constraints are solved
in \cite{eprlett,eprpap} using a master-constraint-like  \cite{master} technique.
In \cite{se}, it was shown that these constraints can equivalently be solved using
a different technique, based on coherent states, yielding the same result.  This
derivation  reinforces  the credibility of the approach, and opens a direct connection
to the semiclassical limit.

In the same paper  \cite{se}, on the other hand, it was also pointed out that
considering a different  class
of coherent  states leads to a variant of the model.  This variant has been
extensively explored in
\cite{fk2007}, and extended to the case of finite $\gamma>1$.  The original model
of  \cite{eprlett,eprpap} and the variant pointed out in  \cite{se} appear as limiting $\gamma\to
0$ and  $\gamma\to\infty$ cases, respectively. All these models (\cite{eprlett,eprpap,se,fk2007},)
are defined by the same vertex, namely the square of the $SU(2)$ Wigner $15j$ symbol; they differ
for the class of boundary states considered and their measure in the spinfoam sum. In
\cite{area_lett}, on the other hand, it was observed that the use of coherent states may not truly
constraint the physical state space of the theory when the constraints are not entirely second
class, and this happens in the limit case $\gamma\to\infty$. Therefore, while the coherent state
technique introduced in \cite{se} appears to work well in the $\gamma\to 0$ case, its
straightforward extension to large $\gamma$ yields a state space larger than the physical state
space of LQG and -one might argue- larger that the proper quantum state space of gravity.
Furthermore, the spectrum of the geometrical operators in this formulation turns out to be quite
different from the standard one of loop quantum gravity \cite{lqg2}.  Here, thus, we reconsider the
finite $\gamma$ case, but we solve the constraints using the same master constraint technique as in
\cite{eprlett,eprpap}. We leave the understanding of our results in terms
of coherent states for future developments.

We find a model with a number of interesting properties.  First, the second class constraints do
reduce the dimension of the physical state space as one wants.  Second, for all values of $\gamma$
the state space precisely matches the one of LQG (on a fixed graph).  This is particularly
interesting in the case of the lorentzian theory,  where such a match is traditionally more
problematic.   Third, the spectrum of the area operator turns out to be discrete, and to be the
same as in LQG, including the correct dependence on the Immirzi parameter $\gamma$.  What is of
particular interest is that this is true in the lorentzian case as well, in spite of the fact that
the unitary representations of the Lorentz group are labelled also by a continuous parameter. This
provides a solution to a long-standing controversy: the area spectrum is discrete in LQG while it
appears to be continuous in the spinfoam framework. The solution is that the area spectrum is
continuous in spinfoams at the kinematical level, but it turns out to become discrete after proper
implementation of the (second class) constraints. Finally, the \emph{ad hoc} ``flip" of the
symplectic structure used to first derive the vertex in \cite{eprlett,eprpap} is not required in
the finite $\gamma$ case.

All these developments rely on two basic ideas.  The first,  championed by Giorgio Immirzi
\cite{Immirzi:1996dr}, is to (``loop") quantize GR by first discretizing it on a Regge-like
triangulation, with appropriately chosen variables. The second is to treat the simplicity
constraints by first imposing them properly in a fixed $SO(4)$ (or $SO(3,1)$) gauge, and then
projecting on the gauge invariant spaces.  The implementation of these ideas is discussed in detail
in \cite{eprlett,eprpap}.  Here, we briefly describe in Section II the discrete theory introduced
in \cite{eprlett,eprpap} (euclidean case) and in \cite{L} (lorentzian case), in order to make this
paper self contained.  We define the euclidean theory in Section III  then the lorentzian theory in
Section IV. Finally, Section 5 is devoted to the spectrum of the area operator.


We work on a fixed triangulation.  It was shown in \cite{rc} that any spinfoam model \cite{SF}, as
the one we define here, admits a group field theory (GFT) \cite{gft} formulation, which is
triangulation independent.  A GFT yielding the new vertex amplitude is indeed already considered in
\cite{fk2007}. We leave the complete construction of the background independent GFT corresponding
to the model defined here to future developments.

\section{Discrete model for finite $\gamma$}

\subsection{Classical theory}

We quantize GR by first discretizing the theory on a Regge geometry.
Introduce a simplicial decomposition $\Delta$ of space-time,
consisting of 4-simplices, tetrahedra, and triangles.  These are
dual respectively to vertices $v$, edges $e$, and faces $f$ in
the dual 2-complex. Geometry is assumed to be flat on each 4-simplex;
curvature is concentrated on the ``bones'' $f$, and is coded in the
holonomy around the ``link'' of each $f$. The variables to describe this
geometry are chosen as follows. (See  \cite{area_lett} for a precise
and detailed definition.)
$e(v)_a^I$ is a tetrad one-form in a cartesian coordinate patch covering
the simplex $v$.  Here $a, b$ are 4d tangent indices and $I,J$ are 4d internal
indices. $e(t)_a^I$ is a tetrad one-form in a cartesian coordinate patch
covering the tetrahedron $t$.  The matrix $(V_{vt})^I{}_J$ is defined
by  $(V_{vt})^I{}_Je(v)_a^J=e(t)_a^I$ when $t$ bounds $v$ and in a
common coordinate patch. It belongs to the group  $G=SO(4)$ in the euclidean case
and $G=SO(3,1)$ in the lorentzian case.

For each triangle $f$ in $t$, define
\begin{equation}
B_f(t)^{IJ} := \int_f \dual(e(t)^I \wedge e(t)^J) .
\end{equation}
where $\dual$ stand for the Hodge dual in the internal  indices. $B_f(t)$
can be seen as element in the algebra $\mathfrak{g}=so(4)$ in the euclidean case
and $\mathfrak{g}=so(3,1)$ in the lorentzian case.
For each triangle $f$ and each pair of tetrahedra $t,t'$ in the link of $f$,
define
\begin{equation}
U_f(t,t') := V_{tv_1} V_{v_1 t_1} V_{t_1 v_2} \cdots V_{v_n t'},
\end{equation}
where the product is around the link in the clock-wise direction
from $t'$ to $t$.

If we  choose $B_f(t)$ as independent variable instead of the tetrads,
the constraints on $B_f(t)$ can be stated as follows.
\begin{itemize}
\item  $\forall$ $f$ and $t,t' \in \Link(f)$,
\begin{equation}U_f(t,t') B_f(t') = B_f(t) U_f(t,t');
\label{1}\end{equation}
\item (closure) $\forall$ $t$
\begin{equation}\sum_{f \in  t} B_f(t) = 0;
\label{2}\end{equation}
\item (diagonal simplicity constraint) $\forall f$
\begin{equation}C_{ff}:=\dual B_f(t)\cdot B_f(t) \approx 0;
\label{3.i}\end{equation}
\item (off-diagonal simplicity constraint)  $\forall f,f' \in t$
\begin{equation}C_{ff'}:=\dual B_f(t)\cdot B_{f'}(t)\approx 0;
\label{3.ii}\end{equation}
\item (dynamical simplicity constraint)  $\forall f,f' \in v$ not in the same $t$
\begin{equation}\dual B_f(v)\cdot B_{f'}(v)\approx \pm 12V(v).
\label{3.iii}\end{equation}
\end{itemize}
The dot stands for the scalar product in the algebra. As noted in
\cite{se,eprpap},  (\ref{3.iii}) is automatically satisfied
when the rest of the constraints are satisfied, due to
the choice of variables; we can therefore forget about it.
When constructing the quantum theory, the above constraints
are incorporated as follows. (\ref{1}) is assumed to hold prior
to varying the action. (\ref{2}) is automatically implemented
by the dynamics in quantum theory, because (\ref{2}) generates
the internal gauge, and the vertex amplitude turns out to project
on the gauge invariant subspace. Finally,  (\ref{3.i})
and (\ref{3.ii}) must be separately imposed on the state space. The
complication is that they are not first class.

The above is the usual formulation of the simplicity constraints. As written, however, the two
constraints (\ref{3.i}) and (\ref{3.ii}) have two sectors of solutions, one in which $B = \dual e
\wedge e$, and one in which $B = e \wedge e$. For finite, non-trivial Immirzi parameter, both
sectors in fact yield GR, but the value of the Newton constant and Immirzi parameter are different
in each sector. In the $B=\dual e \wedge e$ sector, the action \Ref{action} (see below) becomes the
Holst formulation of GR \cite{holst} with Newton constant $G$ and Immirzi parameter $\gamma$. In
the $B=e \wedge e$ sector, one \textit{also} obtains the Holst formulation of GR, but this time
with Newton constants $G\gamma$, and Immirzi parameter $s/\gamma$, where the signature $s$ is +1 in
the euclidean theory and -1 in the lorentzian theory. In order to select a single sector, we
reformulate the simplicity constraints in such a way that these two sectors are distinguished. For
this purpose, we replace (\ref{3.ii}) with
\begin{itemize}
\item For each tetrahedron $t$, {there exists}
an internal vector $n_I$ such that  for all $f \in t$
\be
C_{f}^J:=n_I\ (\dual B_f(t))^{IJ} \approx 0.
\label{3.iib}\ee
\end{itemize}
This condition is stronger than (\ref{3.ii}) since it selects only the desired $B=\dual e \wedge e$
sector. Geometrically, $n_I$ represents of course the normal one-form to the tetrahedron $t$. This
reformulation of the constraint  (\ref{3.ii}) is central to the new models \cite{eprlett, eprpap,
L, se, fk2007}. The vector $n_I$ already played a central role in the covariant loop quantum
gravity approach, which is known to be closely related to 4d spinfoam models \cite{sergei2007,
covariant}.


The classical discrete action is \cite{eprpap}
\begin{eqnarray}
\nonumber
S\!\! &=&\!\! -\frac{1}{2\kappa}\!\! \sum_{f \in int\Delta}\!\! \tr\!\left[B_f(t)U_f(t)
+ \frac{1}{\gamma}\dual B_f(t) U_f(t) \right]  \\
&& -\frac{1}{2\kappa}\!\! \sum_{f \in \partial\Delta}\!\! \tr\!\!\left[B_f(t)U_f(t,t')
+ \frac{1}{\gamma}\dual B_f(t) U_f(t,t') \right]\label{action}
\end{eqnarray}
where $U_f(t):= U_f(t,t)$ is the holonomy around the full link,
starting at $t$, we have set $\kappa = 8 \pi G$, and $int\Delta$ and
$\partial\Delta$ are the interior and the boundary of $\Delta$.
$S$ is a discretization of the continuous action
\begin{eqnarray*}
S &=& \frac{1}{2\kappa} \int_{\scrM} \left[B_{IJ} \wedge F^{IJ}
+ \frac{1}{\gamma} (\dual B)_{IJ} \wedge F^{IJ} \right] \\
&+& \frac{1}{2\kappa} \int_{\partial \scrM} \left[B_{IJ} \wedge F^{IJ}
+ \frac{1}{\gamma} (\dual B)_{IJ} \wedge F^{IJ} \right]
\end{eqnarray*}
which  becomes the Holst action  (see \cite{lqg,holst}) on
substituting in $B=\dual e \wedge e$.

From this we can read off the boundary variables as
$B_f(t) \in \mathfrak{g}$, $U_f(t,t') \in G$.
The variable conjugate to $U_f(t,t')$ is
\begin{equation}
J_f(t) = \frac1\kappa\left(B_f(t) + \frac{1}{\gamma}\dual B_f(t)\right)
\end{equation}
(on the determination of the normalization
factor see \cite{area_lett}.)
That is, the matrix elements $J_f(t)^{IJ}$ have as their
Hamiltonian vector fields the right invariant vector fields
on the group $U_f(t,t')$. Inverting this equation gives
\begin{equation}
\label{Bf}
B_f(t) := \left(\frac{\kappa\gamma^2}{\gamma^2-s}\right)
\left(J_f(t)-\frac{1}{\gamma}\dual J_f(t)\right),
\end{equation}
where $s$ is the signature, namely $s=1$ for $G=SO(4)$
and $s=-1$ for $G=SO(3,1)$. For the cases $\gamma \ll 1$ and $\gamma \gg 1$,
this reduces to
\begin{align}
\nonumber \gamma \ll 1 &\ \leadsto
\  B_f = s\kappa \gamma \dual J_f,  & \gamma \gg 1
&  \ \leadsto\  B_f = \kappa J_f.
\end{align}
corresponding respectively to the flipped and non flipped Poisson
structures of $G$.  The
constraints (\ref{3.i}) and (\ref{3.iib}) can be easily expressed in terms
of the new variables $J_f$
\begin{eqnarray}
C_{ff}&:=&\dual J_f\cdot J_f\left(1+s\frac{1}{\gamma^2}\right)-s\frac{2}{\gamma}J_f\cdot J_f \approx 0 \label{C},\\
C_f^J&:=&n_I\left((\dual J_f)^{IJ}-s\frac{1}{\gamma}J_f^{IJ}\right) \approx 0 \label{CJ}
\end{eqnarray}
(see \cite{eterabf} and \cite{area_lett}) where we have assumed $\gamma$ finite
and $\neq 0,1$.

The closure for the $B_f$ is equivalent to the closure for the
$J_f$ which, as noted above, will be imposed automatically
by the dynamics. In order to proceed, let us fix $n_I=\delta_I^0$. The
general case will be recovered by gauge invariance.
In the lorentzian case this choice restricts all tetrahedra to
be spacelike (it is not clear to us if a non-timelike choice for $n_I$
is viable: see \cite{pr_lorentz31}). With this choice, the constraint \Ref{CJ} becomes:
\begin{equation}
C_f^j=\frac{1}{2}\epsilon^j{}_{kl}J_f^{kl}-s\frac{1}{\gamma}J_f^{0j}=L_f^j-s\frac{1}{\gamma}K_f^j \label{Cj}  \approx 0 ,
\end{equation}
where $\epsilon^{j}{}_{kl}:=\epsilon^{0j}{}_{kl}$,
$L_f^j:=\frac{1}{2}\epsilon^{j}{}_{kl}J_f^{kl}$ are the generators of the $SO(3)$ subgroup
that leaves $n_I$ invariant, and $K_f^j:=J_f^{0j}$ are the generators of the corresponding
boosts.

We take (\ref{C},\ref{Cj}) as our basic set of constraints. So far, we have simply formulated
a discretization of GR.

\subsection{Quantum kinematics}

From the discrete boundary variables and their symplectic
structure, we can write the Hilbert space associated
with a boundary or a 3-slice $\Sigma$.  To do this, it is simpler to
switch to the dual, 2-complex picture, $\Delta^*$.
Let $\Gamma$ be the graph forming the
boundary of $\Delta^*$.  The boundary Hilbert space
is then
\begin{equation}
\Hil = L^2\left(G^{\times L}\right),
\end{equation}
where $L$ is the number of links in $\Gamma$, namely the
number of boundary faces $f$.
As is standard, we replace the groups $G$ with their
universal coverings in the quantum theory.  That is,
from now on $G=Spin(4)$ in the euclidean case
and $G=SL(2,\mathbb{C})$ in the lorentzian case.

Let us concentrate on the single $L^2\left(G\right)$ component
of $\Hil$ associated to a single boundary face $f$. The face $f$ is
dual to the link $l\in\Gamma$ bounding $f$.
The orientation of the triangulation selects one of the
two boundary tetrahedra separated by $f$: call it $t$.
($t$ is dual to the node which is the {\em source} of the link $l$).
For simplicity of notation, let us drop the subscript $f$ and the
dependence $(t)$ all over, and rewrite
$B_f(t)^{IJ}, J_f(t)^{IJ},C^i_f$ simply as $B^{IJ}, J^{IJ}, C^i$.
Let $\hat{J}^{IJ}$ denote the
right-invariant vector fields, determined by the
basis $J^{IJ}$ of $\mathfrak{g}$.
From \Ref{Bf}, the variable $B=B_f(t)$ associated to the
boundary face $f$ (and the $t$ determined by the
orientation mentioned above) is then quantized as
\begin{equation}
\hat{B} := \left(\frac{\kappa\gamma^2}{\gamma^2-s}\right)
\left(\hat{J}-\frac{1}{\gamma}\dual \hat{J}\right).
\end{equation}
Equation \Ref{1} implies then that the quantities associated to the
other tetrahedron $t$ bounding $f$ are given by the
corresponding  {\em left}-invariant vector fields
\cite{eprlett,eprpap}.

Next we impose the constraints \Ref{C} and \Ref{Cj} in the quantum theory.
The constraint \Ref{C} commutes with the others and can be imposed directly
as a strong operator equation. It reads:
\be
C_2\left(1+\frac{s}{\gamma^2}\right)-\frac{2s}{\gamma}C_1\approx 0 \label{diag}
\ee
where $C_1$ and $C_2$ are the Casimir and pseudo-Casimir operators  of $\mathfrak{g}$.
\begin{eqnarray}
C_1&=&J\cdot J=2\left(L^2+sK^2\right), \label{ca1} \\
C_2&=&\dual J\cdot J=4sL\cdot K. \label{ca2}
\end{eqnarray}
$L^2$ is the Casimir of the $SU(2)$ subgroup that leaves $n_I$ fixed. The constraints \Ref{Cj} on
the other hand do not close as a Poisson algebra and their imposition in the quantum theory is more
subtle. We follow the strategy of \cite{master} and replace the set of these constraints with the
single ``master" constraint
\be
M_f:=\sum_i\;(C^i)^2=\sum_i\;\left(L^i-\frac{s}{\gamma}K^i\right)^2 \approx 0.
\ee
which is of course equivalent to the system \Ref{Cj} in the classical theory.
In terms of the Casimir operators, $M_f\approx 0$ gives:
\be
L^2\left(1-\frac{s}{\gamma}\right)+\frac{s}{2\gamma^2}C_1-\frac{1}{2\gamma}C_2\approx 0,
\ee
While equation  \Ref{diag}  was already known \cite{eterabf},
this last relation was noticed only in \cite{area_lett}, as far as we know.
We can finally use \Ref{diag} to simplify this last equation obtaining
\be
C_2=4\gamma L^2.  \label{offdiag}
\ee
The solutions to \Ref{diag}-\Ref{offdiag} will depend on the particular group $G$. In the next sections we analyze separately the cases for $G=Spin(4)$ and $G=SL(2,\mathbb{C})$.

\section{Euclidean theory}

The unitary representation of $Spin(4)$ are labelled by the two half integers $(j^+,j^-)$.
With the usual ordering (and with our normalizations) the two Casimirs (\ref{ca1}) and (\ref{ca2}) have the values
\begin{eqnarray}
C_1&=&4 j^+(j^++1)+ 4j^-(j^-+1), \\
C_2&=&4j^+(j^++1)-4j^-(j^-+1)
\end{eqnarray}
in the representation $(j^+,j^-)$. The constraint \Ref{diag} fixes the ratio between these two
Casimirs in term of the Immirzi parameter $\gamma$. Choosing a suitable ordering (or, equivalently,
up to $\hbar$ corrections), solutions are given by
\be
\label{diageuclid}
(j^+)^2=\left(\frac{\gamma +1}{\gamma -1}\right)^2(j^-)^2.
\ee
An ordering ambiguity is always present in quantum theory. For instance, different orderings of the
Casimir can yields spectra $j(j+1)$ or $j^2$ or $(j+1/2)^2$, or anything differing from these by a
linear or constant shift. In the spinfoam context, this ambiguity can be related to ambiguities in
the path integral measure. In the present context, the natural ordering in order to find solutions
to the simplicity constraints seems to favor the spectrum $j^2$ (or $(j+1/2)^2$) for the $SU(2)$
Casimir operator instead of the usual $j(j+1)$. This would lead to an area spectrum with a constant
spacing of the type $j$ (or $(j+1/2)$).

In \Ref{diageuclid}, we distinguish the two cases: if $\gamma\!>\!0$ then $j^+\!>\!j^-$; while if
$\gamma\!<\!0$ then $j^+\!<\!j^-$. Let us restrict to the case $\gamma\!>\!0$. Notice that this
equation imposes a quantization condition over $\gamma$ as the labels $j^\pm$ are half-integers.
This was pointed out in \cite{fk2007}. Inserting this into the second simplicity constraint
\Ref{offdiag} (and, again, allowing for $\hbar$ corrections) constrains the quantum number $k$
associated to the $SU(2)$ Casimir $L^2$ to
\be
k^2=\left(\frac{2j^-}{1-\gamma}\right)^2 = \left(\frac{2j^+ }{1+\gamma}\right)^2,
\ee
The solutions are substantially different for $\gamma< 1$ and for $\gamma > 1$ (the value
$\gamma=1$ is the natural turning point in the euclidean setting since it corresponds to a
pure self-dual connection):
\be
k=\left\{\begin{array}{cc}j^+ + j^- & \;\; 0<\gamma < 1, \\
                          j^+ - j^- & \;\; \gamma > 1. \\ \end{array}\right.
                          \label{kk}
\ee
That is, for $\gamma <1$, the constraint selects the highest irreducible in the decomposition of
$\mathcal{H}_{(j^+,j^-)}$ when viewed as the carrying space of a reducible representation under the
action of the $SU(2)$ subgroup: $\mathcal{H}_{(j^+,j^-)}=\mathcal{H}_{|j^+-j^-|}\oplus
...\oplus\mathcal{H}_{j^+ +j^-}$. For $\gamma >1$ the lowest irreducible is selected instead.


A prescription similar to ours for the $\gamma < 1$ case already appeared earlier in \cite{fk2007},
where the authors impose $k=j_+ + j_-$  for $\gamma > 1$ for what they call the ``topological"
sector. However the term ``topological" is not accurate; for, what is meant by this is the $B = e
\wedge e$ sector and, as already noted earlier, this sector does not correspond to a topological sector of gravity but truly to
general relativity with effective Immirzi parameter $1/ \gamma$. The model in \cite{fk2007} is
simply related to our $\gamma < 1$ model by $\gamma \mapsto 1/\gamma$. Finally, unlike
\cite{fk2007}, we find solutions to the simplicity constraints in both $\gamma>1$ and $\gamma<1$
cases.

The component of $\Hil=L^2(Spin(4)^{\times L})$ associated to each face $f$ decomposes as
\be
L^2(Spin(4))=\bigoplus_{j^+j^-} \overline{H_{j^+j^-}}\otimes H_{j^+j^-}.
\ee
The diagonal simplicity constraint restricts the direct sum to spins satisfying  \Ref{diageuclid}.
The off-diagonal simplicity constraints select the $SU(2)$ irreducible with the spin determined
by \Ref{kk} in each of the two factors. We call this constrained subspace $\Hil_f$.

$\Hil_f$ can be naturally identified with $L^2(SU(2))$. The projection
\begin{equation}
\pi : L^2\left(Spin(4)\right)\longrightarrow L^2\left(SU(2)\right)  \sim  \Hil_f
\end{equation}
can be written explicitly as follows. A basis in  $L^2(Spin(4))$ is formed by the
matrix elements $D^{(j^+,j^-)}_{q^+q^-,q'^+q'^-}(g)$ of the irreducible
representations.
 Here $g\in Spin(4)$, and the indices $q^\pm$ label a basis in the
representation $j^\pm$. Then
\begin{eqnarray}
\pi\!\!&:&\!\!  D^{(j^+,j^-)}_{q^+q^-,q'^+q'^-}(g)\mapsto
      D^{(j^+,j^-)}_{q^+q^-,q'^+q'^-}(u) \
      c^{{q}^+{q}^-}_{\ m}
      c^{{q'}^+{q'}^-}_{\ m'}.
      \nonumber
      \label{projeucid}
\end{eqnarray}
where  $u\in SU(2)$ and  the
$c^{q^+q^-}_{\ m}$ are the Clebsch-Gordan coefficients that gives the embedding
of the lowest (resp. highest) $SU(2)$ irreducible (where the $m$ index lives) into
the representation $(j^+,j^-)$.
This construction defines also an embedding from the $SU(2)$ spin networks
to the $Spin(4)$ spin networks on $\Gamma$. This is defined by
the embedding of $L^2\left(SU(2)^{\times L}\right)$ into $L^2\left(Spin(4)^{\times L}\right)$
defined by the inclusion $L^2(SU(2))\sim \Hil_f\subset L^2(Spin(4))$
followed by the group averaging over $Spin(4)$ at every node, as
determined by the constraint \Ref{2} (which, we recall, is implemented by
the dynamics).

Let us see how this construction affects the intertwiner spaces.
We decompose the Hilbert space associated with each face into representations.
The simplicity and cross-simplicity constraints, as discussed above,
are then imposed on each of these representations.
Consider four links, colored with the representations  $(j_1^+,j_1^-)...(j_4^+,j_4^-)$,
satisfying \Ref{diageuclid}, meeting at a given node $e$ of $\Gamma$. (This is the
dual picture of four faces bounding a given tetrahedron in the boundary of the
triangulation).
Consider the tensors product of the corresponding representation
spaces $\mathcal{H}_e:=\mathcal{H}_{{\scriptstyle
(j_1^+,j_1^-)}}\otimes ...\otimes \mathcal{H}_{{\scriptstyle
(j_4^+,j_4^-)}}$. Define the constraint $C_e:=\sum_i M_{f_i}$. Imposing
$C_e=0$ strongly on the states in   $\mathcal{H}_0$ selects in each link the
lowest (resp. highest) $SU(2)$ irreducible. Group averaging over
$Spin(4)$ defines then the physical intertwiner
space for the node $e$. The projection from the $Spin(4)$ to the $SU(2)$
intertwiner spaces is then given by:
\begin{eqnarray}
\pi\!\!&:&\!\! Inv_{Spin(4)}\!\left(\mathcal{H}_e\right)\to Inv_{SU(2)}\!\left(\mathcal{H}_{j_1^+ \pm j_1^-}\otimes ...\otimes\mathcal{H}_{j_4^+\pm j_4^-}\right) \nonumber \\
&& C^{(i^+_e,i^-_e)}_{(q_1^+q_1^-)...(q_4^+q_4^-)}\mapsto
C^{(i^+_e,i^-_e)}_{(q_1^+q_1^-)...(q_4^+q_4^-)} \bigotimes_{i=1}^4 c^{q_i^+q_i^-}_{\ m_i}.
\nonumber
\end{eqnarray}
Here $C^{(i^+_e,i^-_e)}_{(q_1^+q_1^-)...(q_4^+,q_4^-)}$ is the normalized intertwiner
defined by a virtual link carrying the $(i^+_e,i^-_e)$ representation.
The corresponding embedding can be written in the form:
\begin{eqnarray}
f\!\!&:&\!\!  Inv_{SU(2)}\left(\mathcal{H}_{k_1}\otimes
...\otimes\mathcal{H}_{k_4}\right)\rightarrow
Inv_{Spin(4)}\left(\mathcal{H}_e\right) \nonumber \\
&& i^{m_1...m_4}\mapsto
\ \ \int_{Spin(4)}\;dg\ \; i^{m_1...m_4} \nonumber
\\
&& \ \ \ \times\  \bigotimes_{i=1}^4\;D^{{\scriptstyle \frac{(1+\gamma)k_i}{2},\frac{|1-\gamma|k_i}{2}}}_{q^+_iq^-_i,q'^+_iq'^-_i} (g) \ \ c^{q'^+_i q'^-_i}_{\ m_i}.
\end{eqnarray}

We are now ready to define the vertex.  For the details of the derivation, see
\cite{plebanski} and \cite{L}.
Following \cite{eprlett, eprpap,L}, the amplitude
of a single vertex bounded by ten $SU(2)$ spins $j_{ab}, a,b=1,...,5$ and five $SU(2)$
intertwiners $i_{a}$ is given by
\begin{eqnarray}
A(j_{ab},i_a)&=&\sum_{i_a^+ i_a^-}
15j\!\left({\scriptstyle \frac{(1+\gamma)j_{ab}}{2}};i_a^+\right)
15j\!\left({\scriptstyle\frac{|1-\gamma|j_{ab}}{2}};i_a^-\right)\nonumber\\
&& \bigotimes_a f^{i_a}_{i_a^+ i_a^-}(j_{ab})
\end{eqnarray}
where the $15j$ are the standard $SU(2)$ Wigner  symbols, and
\be
f^{i}_{i^+ i^-}:=i^{m_1...m_4}
C^{i^+i^-}_{(q_1^+q_1^-)...(q_4^+q_4^-)}
\bigotimes_{i=1...4} c^{q_i^+q_i^-}_{\ m_i}.
\ee
The partition function for an arbitrary triangulation, is
given by gluing these amplitudes together with suitable edge and
face amplitudes. It can be written as:
\be
Z=\sum_{j_f,i_e} \prod_f\;d_f\;\prod_v\;A(j_f,i_e),
\ee
where
\be
d_f:=\left(|1-\gamma|j_f+1\right)\left((1+\gamma)j_f+1\right).
\ee

\section{Lorentzian theory}

The unitary representations in the  principal series are labelled by $(n,\rho)$,
where $n$ is a positive integer and $\rho$ real \cite{lorentz,andersonetal}.  The Casimir
operators for the representation $(n,\rho)$, are given by
\begin{eqnarray}
C_1&=&\frac{1}{2}\left(n^2-\rho^2-4\right), \label{C1} \\
C_2&=&n\rho. \label{C2}
\end{eqnarray}
Up to ordering ambiguities, equation  \Ref{diag} reads now
\be
 n \rho\left(\gamma-\frac{1}{\gamma}\right)=\rho^2- n^2.
 \ee
%
Solutions are given by either $\rho=\gamma n$ or $\rho=-n/\gamma$. The existence of these two
solutions reflects the two sectors mentioned earlier with Immirzi parameter $\gamma$ and
$-1/\gamma$. BF theory can not a priori distinguish between these two sectors (see e.g.
\cite{eterabf}). However, in our framework, the second constraint \Ref{offdiag} breaks this symmetry and select the
first branch $\rho=\gamma n$. It further imposes that $k=n/2$, where $k$ again labels the subspaces
diagonalizing $L^2$. Therefore the constraints select the lowest $SU(2)$ irreducible representation
in the decomposition of $\mathcal{H}_{(n,\rho)}=\bigoplus_{k\geq n/2}\mathcal{H}_k$. This choice of
the lowest weight corresponds to the usual notion of coherent states for the non-compact $SL(2,\C)$
Lie group \cite{Perelomov} (see also \cite{eteradaniele}). Notice that there is restriction on the
value of $\gamma$ as there was in the Euclidean case.

Notice also that the continuous label $\rho$ becomes quantized, because $n$ is discrete. It is
because of this fact that any continuous spectrum depending on $\rho$ comes out effectively
discrete on the subspace satisfying the simplicity constraints.

This construction defines the projection from
the $SL(2,\mathbb{C})$ boundary Hilbert space to the $SU(2)$ space.
For a single $D$ matrix, this projection reads (see the \cite{L}):
\begin{eqnarray}
\pi\; &:&\; L^2\left(SL(2,\mathbb{C})\right)\longrightarrow L^2\left(SU(2)\right) \nonumber \\
      &&D^{n,\rho}_{jqj'q'}(g)\longmapsto D^{n/2}_{qq'}(u) \label{projlorentz}
\end{eqnarray}
This also
defines an embedding from the $SU(2)$ Hilbert space to the $SL(2,\mathbb{C})$ space, given by inclusion followed by group
averaging over the Lorentz group.

As before, in order to extend this result to the complete space $\mathcal{H}$ we have to define the
projection for the intertwiners. Consider four links meeting at a given node $e$ of $\Gamma$,
carrying representations $(n_1,\rho_1)...(n_4,\rho_4)$, satisfying the diagonal constraints.
Consider the Hilbert space of tensors between these representations:
$\mathcal{H}_e:=\mathcal{H}_{{\scriptstyle (n_1,\rho_1)}}\otimes ...\otimes
\mathcal{H}_{{\scriptstyle (n_4,\rho_4)}}$. Construct the constraint $C_e:=\sum_i M_{f_i}$.
Imposing $C_e=0$ strongly selects in each link the lowest $SU(2)$ along with the representations of
the form $\rho=n\gamma$. The last step is group averaging over $SL(2,\mathbb{C})$ which defines the
physical intertwiner space for this node. The projection is then given by:
\begin{eqnarray}
\pi\!\! &:&\!\! Inv_{SL(2,\mathbb{C})}\left(\mathcal{H}_e\right)\longrightarrow
Inv_{SU(2)}\left(\mathcal{H}_{\frac{n_1}{2}}\otimes ...\otimes\mathcal{H}_{\frac{n_4}{2}}\right) ,
\nonumber \\
&& C^{(n_e,\rho_e)}_{(j_1,q_1)...(j_4,q_4)}\longmapsto
C^{(n_e,\rho_e)}_{(\frac{n_1}{2},q_1),...(\frac{n_4}{2},q_4)}.
\end{eqnarray}
The embedding is given by:
\begin{eqnarray}
f\!\! &:&\!\! Inv_{SU(2)}\left(\mathcal{H}_{j_1}\otimes
...\otimes\mathcal{H}_{j_4}\right)\longrightarrow  Inv_{SL(2,\mathbb{C})}\left(\mathcal{H}_e\right), \nonumber \\
&&\!\! i^{m_1...m_4}\longmapsto\int_{SL(2,\mathbb{C})}\!\!\!\!dg\  i^{m_1...m_4}
\bigotimes_{i=1}^{i=4}\;D^{(2j_i,2j_i\gamma)}_{(j'_i,m'_i)(j_i,m_i)}(g).\nonumber
\end{eqnarray}
The boundary space is once again just given by the $SU(2)$ spin networks.

We are now ready to define the vertex. As before, we obtain
\begin{eqnarray}
A(j_{ab},i_a)&=&\sum_{n_a}\int d\rho_a (n_a^2+\rho_a^2)
\left(\bigotimes_a\
f^{i_a}_{n_a\rho_a}(j_{ab})\right)\nonumber \\
&&
 15j_{SL(2,\mathbb{C})}\left((2j_{ab},2j_{ab}\gamma);(n_a,\rho_a)\right)\
\end{eqnarray}
where we are now using the 15j of $SL(2,\mathbb{C})$ and
\be
f^{i}_{n\rho}:=i^{m_1...m_4}\ \bar{C}^{n\rho}_{(j_1,m_1)...(j_4,m_4)},
\ee
where $j_1...j_4$ are the representations meeting at the node.
The final partition function, for an arbitrary triangulation, is
given by gluing these amplitudes together with suitable edge and
face amplitudes:
\be
Z=\sum_{j_f,i_e} \prod_f\;(2j_f)^2(1+\gamma^2)\;\prod_v\;A(j_f,i_e).
\ee

\section{Area spectra}

There are two operators related to the area of a triangle dual to the face $f$.
\be
A_4(f):=\frac12(\dual B)^{IJ}(\dual B)_{IJ}
\ee
and its projected (gauge fixed) counterpart:
\be
A_3(f):=\frac12(\dual B)^{ij}(\dual B)_{ij}
\ee
Classically, these two quantities are equal due to the constraint \Ref{CJ}. After quantization this
will not hold anymore. This can be seen as follows.  Since boosts do not commute, it is not
possible in the quantum theory to physically implement a Lorentz frame exactly. Hence all spacelike
vectors are affected by quantum fluctuation in the timelike directions. The relation between the
two quantities above is given by
\be
A_4=A_3+\left(\frac{\kappa\gamma^2}{\gamma^2-s}\right)^2  sM_f.
\ee
Let us focus on $A_3$, which is the standard canonical operator considered in a
canonical quantization of GR. We can write
\be
A_3=\left(\frac{\kappa\gamma^2}{\gamma^2-s}\right)^2\left(\vec{K}-\frac{\vec{L}}
{\gamma}\right)^2.
\ee
Using the constraints \Ref{diag} and \Ref{offdiag}, we get with straightforward algebra
\be
A_3=\kappa^2\gamma^2L^2
\ee
for both euclidean and lorentzian signatures. The spectrum is therefore
\be
Area =  \sqrt{A_3} = 8\pi\hbar G\,  \gamma \,  \sqrt{k(k+1)}.
\label{lqgs}
\ee
which is {\em exactly} the spectrum of LQG.
This spectrum can be compared with the continuous spectrum
\be
Area \sim \frac12 \sqrt{4k(k+1)-n^2+\rho^2+4}.
\label{cons}
\ee
that was previously obtained in covariant LQG, before imposing the second class
constraints (see \cite{covariant}). Remarkably, imposing the simplicity
constraints \Ref{diag} and \Ref{offdiag} reduces the continuous spectrum
\Ref{cons} to the exact discrete LQG spectrum \Ref{lqgs}.

Finally, we would like to point out that the ordering of the Casimir operators for $SU(2)$ and
$SL(2,\C)$ required to have meaningful simplicity constraints do not use the usual ordering but
seems to select an area spectrum with a regular spacing such as $j$ (or $j+1/2$) instead of the
standard $\sqrt{j(j+1)}$. This issue deserves further investigation.

\section{Conclusion}

We have defined a spinfoam model for finite values of the Immirzi parameter
$\gamma$, for the euclidean as well as for the lorentzian theory.  In both
cases, the boundary
space turns out to be the same as in LQG, spanned by $SU(2)$ spin networks. The
spectrum of the area operator too is the same as in LQG, both for the euclidean
and the lorentzian sectors.

We leave the analysis of the model for future developments.
Among the numerous issues we leave open is whether the vertex itself is finite
in the lorentzian case, or whether it needs to be regulated; and whether it is
possible to extend the spinfoam finiteness results \cite{finiteness} to the present
model.  It would be of particular interest to check whether this model gives the
correct graviton propagator \cite{propagator}.

One of the main results of this paper is to bring LQG and the spinfoam
formalism much closer, in four dimensions.  It would be be of great interest
if a direct relation between these two nonperturbative quantizations of GR
could be completely established, as it was done in three dimensions by
Perez and Noui \cite{PN}.   For this, it would be necessary to write
the hamiltonian constraint operator that generates the new vertex, that is,
whose matrix elements are given by the new vertex amplitude.

\centerline{---------}

Thanks to Alejandro Perez, Simone Speziale and Laurent Freidel for numerous exchanges.


\begin{thebibliography}{99}

\bibitem{BC}
JW Barrett, L Crane, ``A lorentzian signature model for quantum GR'', \textit{Class
Quant Grav} \textbf{17} (2000) 3101-3118.
JW Barrett, L Crane, ``Relativistic spin networks and quantum
gravity'', \textit{J Math Phys} \textbf{39} (1998) 3296-3302.
R DePietri, L Freidel, K Krasnov, C Rovelli ``Barrett-Crane
model from a Boulatov-Ooguri field theory over a homogeneous space",
\textit{Nucl Phys} B\textbf{574} (2000) 785-806.
A Perez, C Rovelli, ``A spinfoam model without bubble divergences",
\textit{Nucl Phys} B\textbf{599} (2001) 255-282.
D Oriti, RM Williams, ``Gluing 4-simplices: a derivation of the
Barrett-Crane spinfoam model for Euclidean quantum gravity",
\textit{Phys Rev} \textbf{D63}(2001) 024022.


\bibitem{eprlett}
J Engle, R Pereira, and C Rovelli ``The loop-quantum-gravity
vertex-amplitude'', \textit{Phys Rev Lett} \textbf{99} 161301
(2007).

\bibitem{eprpap}
J Engle, R Pereira, and C Rovelli, ``Flipped spinfoam vertex and
loop gravity'', \texttt{arXiv:0708.1236}.

\bibitem{ema} E Alesci, C Rovelli, ``The complete LQG graviton propagator:
I. difficulties with the Barrett-Crane vertex'', \texttt{arXiv:0708.0883}.

\bibitem{numer}
E Magliaro, C Perini, C Rovelli, ``Numerical indications on the semiclassical limit of the flipped vertex", arXiv:0710.5034.

\bibitem{lqg}
  A Ashtekar,
  An introduction to loop quantum gravity through cosmology,
  gr-qc/0702030.
  A Ashtekar, J Lewandowski,
  Background independent quantum gravity: A status report,
 {\em Class Quant Grav} {\bf 21} (2004) R53-R152.
 T Thiemann, \emph{Introduction to Modern Canonical Quantum General Relativity}
(CUP, 2007).
 C Rovelli, \emph{Quantum Gravity} (CUP, Cambridge, 2004).
 C Rovelli, L Smolin,
  ``Knot theory and quantum gravity''
 \textit{Phys  Rev  Lett } \textbf{61} (1988) 1155-1158. 
 C Rovelli, L Smolin,
 ``Loop space representation for quantum GR",
 \textit{Nucl Phys} \textbf{B331}  (1990) 80-152. 
 A Ashtekar, C Rovelli, L Smolin,
 ``Weaving a classical geometry with quantum threads",
 \textit{Phys Rev Lett}  {\bf 69}, 237 (1992).

\bibitem{L}
R Pereira,  ``Lorentzian LQG vertex amplitude",  arXiv:0710.5043.

\bibitem{sergei2007}
S Alexandrov, ``Choice of connection in Loop Quantum Gravity'' \textit{Phys  Rev}  {\bf D65}
(2002) 024011.
S Alexandrov, E Livine, ``SU(2) Loop Quantum Gravity seen from
Covariant Theory'', \textit{Phys Rev}  {\bf D67} (2003) 044009.
E Buffenoir, M Henneaux, K Noui, Ph Roche,
``Hamiltonian Analysis of Plebanski Theory", {\it Class Quant Grav} {\bf 21} (2004) 5203-5220.
S Alexandrov, ``Spin foam model from canonical quantization'', \texttt{arXiv:0705.3892}.

\bibitem{covariant}
 E Livine,  ``Towards a covariant loop quantum gravity'',
in {\em Approaches to quantum gravity},  D Oriti  ed (Cambridge University Press,
to appear),  arXiv:gr-qc/0608135.

\bibitem{master}
T Thiemann,   ``The Phoenix Project: Master Constraint Programme for Loop Quantum Gravity"
\textit{Class  Quant  Grav }   \textbf{23} (2006)  2211-2248, gr-qc/0305080.
B Dittrich, T Thiemann
``Testing the Master Constraint Programme for Loop Quantum Gravity I. General Framework"
\textit{Class  Quant  Grav }  \textbf{23} (2006)  1025-1066, gr-qc/04011138,
B Dittrich, T Thiemann
``Testing the Master Constraint Programme for Loop Quantum Gravity II. Finite Dimensional Systems"
\textit{Class  Quant  Grav }  \textbf{23} (2006)  1067-1088, gr-qc/0411139.
T Thiemann,
``Quantum spin dynamics VIII. The Master Constraint"
\textit{Class  Quant  Grav } \textbf{23} (2006) 2249-2266.


\bibitem{se}
E Livine, S Speziale ``A new spinfoam vertex for quantum gravity'', \texttt{arXiv:0705.0674}.
E Livine, S Speziale ``Consistently solving the simplicity constraints for spinfoam quantum
gravity'', \texttt{arXiv:0708.1915}.

\bibitem{fk2007}
L Freidel, K Krasnov, ``A new spin foam model for 4d gravity'',
\texttt{arXiv:0708.1595}.

\bibitem{area_lett}
J Engle, R Pereira, ``Coherent states, constraint classes, and area operators in the new spin-foam models'', arXiv:0710.5017.

\bibitem{lqg2}
C Rovelli, L Smolin,
 ``Discreteness of Area and Volume in Quantum Gravity",
\textit{Nucl Phys} \textbf{B442} (1995)  593-619;
   \textit{Nucl Phys} \textbf{B456}, 734 (1995).
A Ashtekar, J Lewandowski,
``Quantum Theory of Geometry I: Area Operators"
{\em Class  Quant Grav} {14}
(1997) A55-A82;
``II: Volume Operators",
{\em Adv Theo Math Phys} {1} (1997) 388-429.

\bibitem{Immirzi:1996dr}
  G Immirzi,
  ``Quantum gravity and Regge calculus",
   {\it Nucl Phys Proc Suppl}  {\bf 57} (1997) 65,
  arXiv:gr-qc/9701052.

\bibitem{rc}
M Reisenberger, C Rovelli,
 ``Spacetime as a Feynman diagram: The connection formulation",
  \textit{Class Quant Grav} {\bf 18} (2001)  121-140. 
M Reisenberger, C Rovelli,
 ``Spin foams as Feynman diagrams''
  in \emph{Florence 2001, A relativistic spacetime odyssey} ed.
  I Ciufolini, D Dominici, L Lusanna
  (World Scientific, Singapore, 2003) pg 431-448.

\bibitem{SF}
D Oriti, ``Spacetime geometry from algebra: spin foam models for
non-perturbative quantum gravity'', \textit{Rept Prog Phys}
\textbf{64} (2001) 1489-1544.
A Perez, ``Spin Foam Models for Quantum Gravity", \emph{Class
Quantum Grav} \textbf{20} (2003) R43.
 MP Reisenberger, C Rovelli,
  `` `Sum over surfaces' form of loop quantum gravity,''
  \textit{Phys Rev}  {\bf D56}, 3490 (1997).
   MP Reisenberger,
  ``A lattice worldsheet sum for 4-d euclidean GR'',
  arXiv:gr-qc/9711052.
JC Baez, ``Spin foam models,''
 \textit{Class Quant Grav}  {\bf 15} (1998) 1827-1858.
JC Baez,  ``An introduction to spin foam models of BF theory and quantum gravity,''
 \textit{Lect Notes Phys}  {\bf 543} (2000) 25-94.

\bibitem{gft}
  D Oriti,
  ``The group field theory approach to quantum gravity,''
  arXiv:gr-qc/0607032.
    L Freidel,  ``Group Field Theory: an overview'',
    \textit{Int Journ Theor Phys} {\bf 44} (2005) 1769-1783.
    R DePietri, L Freidel, K Krasnov, C Rovelli ``Barrett-Crane
    model from a Boulatov-Ooguri field theory over a homogeneous space",
    \textit{Nucl Phys} \textbf{B574} (2000) 785-806.

\bibitem{holst} S Holst,  ``Barbero's Hamiltonian derived from a
generalized Hilbert-Palatini action",
\textit{Phys Rev} \textbf{D53} (1996) 5966-5969.

\bibitem{eterabf}
  RE Livine, D Oriti,
  ``Barrett-Crane spin foam model from generalized BF type action for gravity'',
  {\it Phys Rev } {\bf D65} (2002) 044025.

\bibitem{pr_lorentz31} A Perez, C Rovelli, ``3+1 spinfoam model of quantum gravity with spacelike and timelike components'',  \textit{Phys Rev} \textbf{D64} (2001) 064002.
\texttt{arXiv:0011037}.

\bibitem{lorentz}
W Ruhl, \emph{The Lorentz group and harmonic analysis} (WA Benjamin Inc, New York, 1970).
MA Naimark, \emph{Les repr\'esentations lin\'eaires du groupe de Lorentz} (Dunod, Paris, 1962).
IM Gel'fand, MI Graev, NYa Vilenkin, \emph{Generalized Functions: Volume 5 Integral Geometry and Representation Theory} (Academic Press, 1966).

\bibitem{andersonetal} RL Anderson, R Raczka, MA Rashid, P Winternitz, ``Clebsch-Gordan coefficients for the Lorentz group - I: Principal Series'', ICTP, Trieste, IC/67/50, (1967).

\bibitem{plebanski}
  A Perez,
  ``Spin foam quantization of SO(4) Plebanski's action'',
 \textit{Adv Theor Math Phys} {\bf 5} (2002) 947
  [Erratum-ibid.\,{\bf 6} (2003) 593].

\bibitem{Perelomov} AM Perelomov, {\em Generalized Coherent States and Their Applications}
(Springer-Verlag, Berlin 1986).

\bibitem{eteradaniele}
  RE Livine, D Oriti,
  ``Coherent States for 3d Deformed Special Relativity: semi-classical points in a quantum flat spacetime'',   {\it JHEP } (2005) 050.

\bibitem{finiteness} A Perez,
``Finiteness of a spinfoam model for euclidean GR'',
 \textit{Nucl Phys}  \textbf{B599} (2001) 427-434.
L Crane, A Perez, C Rovelli,
 ``Finiteness in spinfoam quantum gravity",
\textit{Phys  Rev  Lett } \textbf{87} (2001) 181301.
A Perez, C Rovelli,
{\em Phys Rev} {\bf D63} (2001) 041501.
  ``Spin foam model for lorentzian GR,''
L Crane, A Perez, C Rovelli,
``A finiteness proof for the lorentzian state sum spinfoam model for  quantum GR",
gr-qc/0104057.

\bibitem{propagator}
C Rovelli, ``Graviton propagator from
background--independent quantum gravity'',
 \textit{Phys Rev Lett}   \textbf{97}  (2006) 151301,
 arXiv:gr-qc/0508124.
L Modesto, C Rovelli: ``Particle scattering in loop
quantum gravity",
\textit{Phys Rev Lett}   \textbf{95}   (2005) 191301.
E Bianchi, L Modesto, C Rovelli, S Speziale
``Graviton propagator in loop quantum gravity''
\textit{Class Quant Grav} \textbf{23} (2006) 6989-7028.
S Speziale, ``Towards the graviton from spinfoams:
the 3d toy model",
\textit{JHEP} {\bf 0605}, 039 (2006),
  arXiv:gr-qc/0512102.
E Livine, S Speziale, J Willis, ``Towards the graviton
from spinfoams:
higher order corrections in the 3d toy model",
{\em Phys Rev}   {\bf D75}, 024038 (2007),
arXiv:gr-qc/0605123.
  ER Livine, S Speziale,
``Group integral techniques for the spinfoam graviton propagator'',
{\em  JHEP} {\bf 0611} (2006) 092.
 B Dittrich, L Freidel, S Speziale,
  ``Linearized dynamics from the 4-simplex Regge action'',
  arXiv:0707.4513.
E Bianchi, L Modesto,
  ``The perturbative Regge-calculus regime of Loop Quantum Gravity'',
  arXiv:0709.2051.

\bibitem{PN} K Noui,  A Perez,
 ``Three dimensional loop quantum gravity: physical scalar product and spin foam models'',
\textit{Class Quant Grav} \textbf{22} (2005) 1739-1762.


\end{thebibliography}
\end{document}